# Stochastic Gradient Descent in the Optimal Control of Execution Costs

Quac Tran-Dinh & Simeon Kolev

RTG: Foundations in Probability, Optimization and Data Sciences

UNC Spring 2024

# Abstract

Bertsimas and Lo's seminal work laid the groundwork for addressing the implementation shortfall dilemma in institutional investing, emphasizing the significance of market microstructure and price dynamics in minimizing execution costs. However, the ability to derive a theoretical Optimum market order policy is an unrealistic assumption for many investors. This study aims to bridge this gap by proposing an approach that leverages stochastic gradient descent (SGD) to derive alternative solutions for optimizing execution cost policies in dynamic markets where explicit mathematical solutions may not yet exist. The proposed methodology assumes the existence of a mathematically derived optimal solution that is a function of the underlying market dynamics. By iteratively refining strategies using SGD, economists can adapt their approaches over time based on evolving execution strategies. While these SGD-based solutions may not achieve optimality, they offer valuable insights into optimizing policies under complex market frameworks. These results serve as a bridge for economists and mathematicians, facilitating the study of the Optimum policy volatile markets while offering SGD driven implementable policies that closely approximate optimal outcomes within shorter time frames.

# Contents

# Introduction

Large Institutional Investor face particular problems when it comes to the measurement and management of profits and trading costs. The incomprehensible amount of assets that mutual, pension, and hedge funds manage gives them the privilege of reliable sustenance at a risk-free rate of returns. For this reason, it is often, that large asset managers prioritize efficient trade execution to keep as much of their earnings as possible. Many large firms have found that a shift in emphasis in the management execution costs (commission, bid/ask spread, and order size costs) may have a more significant impact on retained earnings than strategies prioritizing marginal improvements in trade yields. Therefore, it is important for investment analysts to properly account for the profitability of trades and order size related costs when deriving "Optimal" strategies. However, this can be a difficult obstacle to overcome, and many asset managers have found that "implementation shortfall" is a surprisingly large hinderance to portfolio growth, underscoring the importance of the optimal control of execution costs.

In this paper, I will be exploring the additive price-impact models financial models proposed in the early 2000s that aim to minimize implementation shortfall. Using established and tested methods I will be deriving the "Optimal Strategy" as proposed by the literature. The original optimum strategy is an explicit function of the market parameters, solved using dynamic stochastic programming of a linear optimization problem. Using the optimization techniques from STOR 512 I will derive alternative strategies using Stochastic Gradient Descent. The 4 techniques are Adaptive Gradient Algorithm, Root Mean Square Propagation, Adaptive Moment Estimation, and my own Custom SGD technique. After properly tuning the SGD algorithms, I will compare each model's predicted optimum purchasing order size, the total accumulated execution cost, and the per period variance in the SGD's strategy vs the Optimum strategy.

## Literature Review

The "Optimal Control of Execution Costs" by Dimitris Bertsimas and Andrew W. Lo (1998). at MIT laid the foundation for many price impact models that aim to mitigate excess execution costs. These models serve as a template for equity market conditions and price dynamics. The objective when mitigating implementation shortfall is to make the best trades given the nature of current and future market dynamics as described by the models. Defining and controlling executions costs are fundamentally dynamic optimization problems and have previously been must be solved as such. To optimize execution costs, the dynamic strategy must take into consideration current and future prices and execute trades in such a manner as to minimize total expected cost. The strategies derived from these models aim to exploit market dynamics and minimize the expected costs of executing large trades over a fixed horizon. Specifically, given a fixed block of shares ($S_t$) to be executed within a fixed finite number of periods ($T$).

The general price impact models analyzed by Bertsimas and Lo are additive permanent price impact models. The price models serve as discrete approximations of a stock's price ($P_t$) given, order size ($B_t$), exogenous variables ($X_t$), time-series correlations ($\rho$), parameter relevance ($\theta, \gamma$), and random fluctuations ($\varepsilon_t, \eta_t$). The optimal ($B_t$) strategies are derived by minimizing the Bellman equation ($\sum_1^T P_t B_t$) and solving recursively through stochastic dynamic programing to obtain an explicit closed form expression of the "best-execution strategy."

The following equations summarizes the Additive Permanent Price Impact with Information:

$$P_t = P_{t-1} + \theta B_t + \gamma X_t + \varepsilon_t, \qquad X_t = \rho X_{t-1} + \eta_t$$

$$\varepsilon_t \sim N(0, \sigma_\varepsilon^2), \qquad \eta_t \sim N(0, \sigma_\eta^2)$$

The objective is to find the $B_t$ producing the minimal execution cost strategy given by the Bellman equation, and dynamic parameters:

$$V_t = \min_{B_t} \sum_t^T P_t B_t$$

$$e_t = \frac{1}{T-t+1}, \quad f_t = \frac{\gamma}{\theta(T-t+1)} \sum_1^t (t-k)\rho^k$$

Solving the Bellman equation by dynamic linear programming yields the optimal order size:

$$B_t = e_t S_t + f_t X_{t-1}$$

$$0 \leq B_t \leq S_t, \quad \frac{1}{T-t+1} \sum_1^T B_t = S_t$$

Figure 1 shows the simulated total execution cost under the Optimum Strategy:

```
Optimal Strategy
Strategy Execution Costs 5267079.741349543
Standard Deviation per Period 55.93446146124703
```

*Figure 1*

## Methods of Stochastic Gradient Descent

This section explains the methodology of 4 SGD techniques to derive similar solutions for the optimal $B_t$ and total execution cost as found by Bertsimas and Lo. All SGD techniques are intuitively calibrated with the following initial parameters:

$$\eta = 0.025, \quad \sigma_\varepsilon = 0.125, \quad \beta_1 = 0.98, \quad \beta_2 = 0.99, \quad \max i = 10{,}000,$$

$\eta$ chosen such that $\nabla B_{1,t} = 1$ share, $\sigma_\varepsilon$ chosen proportional to the stock's variance ($\$0.125$ or 1 stock tick), $\beta_1$ and $\beta_2$ chosen such that $\nabla \eta_i$ is sufficient for all $i$, and $\max i$ sufficiently large to allow for relative convergence.

All SGD techniques navigated the box constraints in each period ($t$) the following ways:

1.) $0 \leq B_{i,t} \leq S_t$, $\quad B_{i+1,t} = \begin{cases} 0, & B_{i,t} \leq 0 \\ B_{i,t} + \eta \times \nabla V_{i,t}, & 0 \leq B_{i,t} \leq S_t \\ S_t, & B_{i,t} \geq S_t \end{cases}$

2.) $\frac{1}{T-t+1}\sum_1^t B_{i,k} = S_t$, $\quad B_{i+1,t} = B_{i,t} \times \frac{S_t - \sum_1^t B_{i,k}}{\sum_t^T B_{i,k}}$

### Adaptive Gradient Algorithm

Adaptive Gradient Algorithm (AdaGrad) is a modification to traditional SGD that adapts the learning rate ($\eta$) based on the gradient of the objective function ($\nabla V_t$). AdaGrad performs smaller updates for $B_t$, associated with frequently occurring $P_t$, and conversely larger updates associated with infrequently occurring $P_t$. AdaGrad updates $\eta$ at each iteration ($i$) for every $B_{i,t}$ based on the cumulative sum of previous gradients $\nabla V_{i,t}{}^2$.

$$G_{i,t} = G_{i-1,t} + \nabla V_{i,t}^{2}, \quad B_{i+1,t} = B_{i,t} - \frac{\eta}{\sqrt{G_{i,t} + \epsilon}} \nabla V_{i,t}$$

AdaGrad is very effective at adapting the learning rate based on the historical gradients which can improve convergence and training stability as the model. AdaGrad is also very effective with a high dimensional features space, however as $\sum_{0}^{i}(\nabla V_{k-1,t})^2 \to \infty$, the technique can struggle to make continuously significant improvements as the algorithm happens to produce overly aggressive η reductions.

Figure 2 shows the simulated total execution cost under the AdaGrad Strategy:

```
AdaGrad Strategy
Strategy Execution Costs 5268158.4111539135
Excess Cost per Share 0.010786698043700308
Increase in Standard Deviation per Share 55.253331162117135
```

*Figure 2*

### Root Mean Square Propagation

Root Mean Square Propagation (RMSprop) is an adaptive learning rate SGD algorithm designed to address the issues of NN training. RMSprop adapts the learning rate (η) based on a moving average of the squared gradient of the objective function ($\nabla V_{i,t}^{2}$). RMSprop ensures that descent oscillations are dampened in the direction of steep $\nabla V_t$ and increased in the direction of flat regions of the loss function.

$$G_{i,t} = \beta_1 G_{i-1,t} + (1 - \beta_1)\nabla V_{i,t}^{2}, \quad B_{i+1,t} = B_{i,t} - \frac{\eta}{\sqrt{G_{i,t} + \epsilon}} \nabla V_{i,t}$$

RMSprop is very effective at handing situations where parameters have different scales, and the loss landscape is dynamic over $t$. The technique is very potent as it was adapted for NN which typically solves very complicated dynamic problems; however, it can struggle with overcoming saddle points and requires careful tuning of the decay hyperparameter ($\beta$).

Figure 3 shows the simulated total execution cost under the RMSprop Strategy:

```
RMSprop Strategy
Strategy Execution Costs 5269090.608715242
Excess Cost per Share 0.0201086736569833
Increase in Standard Deviation per Share 81.43011339113211
```

*Figure 3*

## Adaptive Moment Estimation

Adaptive Moment Estimation (Adam) is an adaptive learning rate SGD algorithm for NN optimization that incorporates corrections terms $(\widehat{m_{i,t}}, \widehat{v_{i,t}})$ to account for bias in early iterations. Adam builds on previous learning rate adjustments of by adapting the learning rate ($\eta$) of individual $B_t$ based on the historical gradients and exponentially decaying moving averages of past $\nabla V_{i,t}$ and $\nabla V_{i,t}^2$.

$$m_{i,t} = \beta_1 m_{i-1,t} + (1-\beta_1)\nabla V_{i,t}, \qquad v_{i,t} = \beta_2 v_{i-1,t} + (1-\beta_2)\nabla V_{i,t}^2$$

$$\widehat{m_{i,t}} = \frac{m_{i,t}}{1-\beta_1^i}, \qquad \widehat{v_{i,t}} = \frac{v_{i,t}}{1-\beta_2^i}$$

$$B_{i+1,t} = B_{i,t} - \frac{\eta}{\sqrt{\widehat{v_{i,t}}+\epsilon}} \widehat{m_{i,t}}$$

Adam is effective at adapting the learning rates based on the historical gradients, which can improve convergence and stability, however, it can lead to overly adaptive η and noisy $\nabla V_t$. Adam is a versatile and widely used optimization algorithm that is suitable for various NN problems: however, it can struggle with overcoming saddle points and requires careful tuning of the decay hyperparameters, $(\beta_1, \beta_2)$.

Figure 4 shows the simulated total execution cost under the RMSprop Strategy:

```
Adam Strategy
Strategy Execution Costs 5269085.709459347
Excess Cost per Share 0.020059681098032744
Increase in Standard Deviation per Share 81.30524170826655
```

*Figure 4*

### Custom SGD

My Custom SGD algorithm is a novel modification to the traditional SGD technique. I incorporate an adaptive learning rate (η) and adaptive iterations ($\max i$) such that if the progressively weighted average of the gradient's norm of the previous iterations is less than the current gradients norm, then to decrease η and increase iterations, and conversely to increase η and decrease iterations.

$$f(\eta_i, max\ i) = \begin{cases} \left(\eta_{i-1} + \dfrac{2}{\max i},\ max\ i + \dfrac{0.5}{\max i}\right), & \dfrac{1}{10}\sum_{i-10}^{i} k\|\nabla V_{k,t}\| \geq \|\nabla V_{i,t}\| \\ \left(\eta_{i-1} + \dfrac{0.5}{\max i},\ max\ i + \dfrac{2}{\max i}\right), & \dfrac{1}{10}\sum_{i-10}^{i} k\|\nabla V_{k,t}\| \leq \|\nabla V_{i,t}\| \end{cases}$$

Additionally, the box constraint for the optimal purchasing strategy ($0 \leq B_{i,t} \leq S_t$) was changed such that if $B_{i,t}$ falls outside this range, then $B_{i+1,t} = \frac{S_t}{T-t+1}$, essentially producing a uniform split of the remaining outstanding shares, and resetting the SGD's initialization for period $t$ up to interation $i$ to avoid extreme solutions of $B_{i,t}$.

Figure 5 shows the simulated total execution cost under the RMSprop Strategy:

```
Custom Strategy
Strategy Execution Costs 5268126.133522306
Excess Cost per Share 0.010463921727621927
Increase in Standard Deviation per Period 54.187645165456935
```

*Figure 5*

# Strategy & Analysis

The Optimum and SGD strategies were simulated under the same sequences of $(\varepsilon, \eta)$ producing identically random fluctuations in the stock price $(P_t)$ and serially correlated market information $(X_t)$. Though the actual implemented stock purchasing strategy $(B_t)$ differs drastically between each SGD technique, the optimal minimized accumulated execution cost is practically identical. However, the per period standard deviation of the Optimum $B_t$ is significantly lower than that of the SGD strategies.

Figure 6 shows each strategy's simulated per period buying strategy:

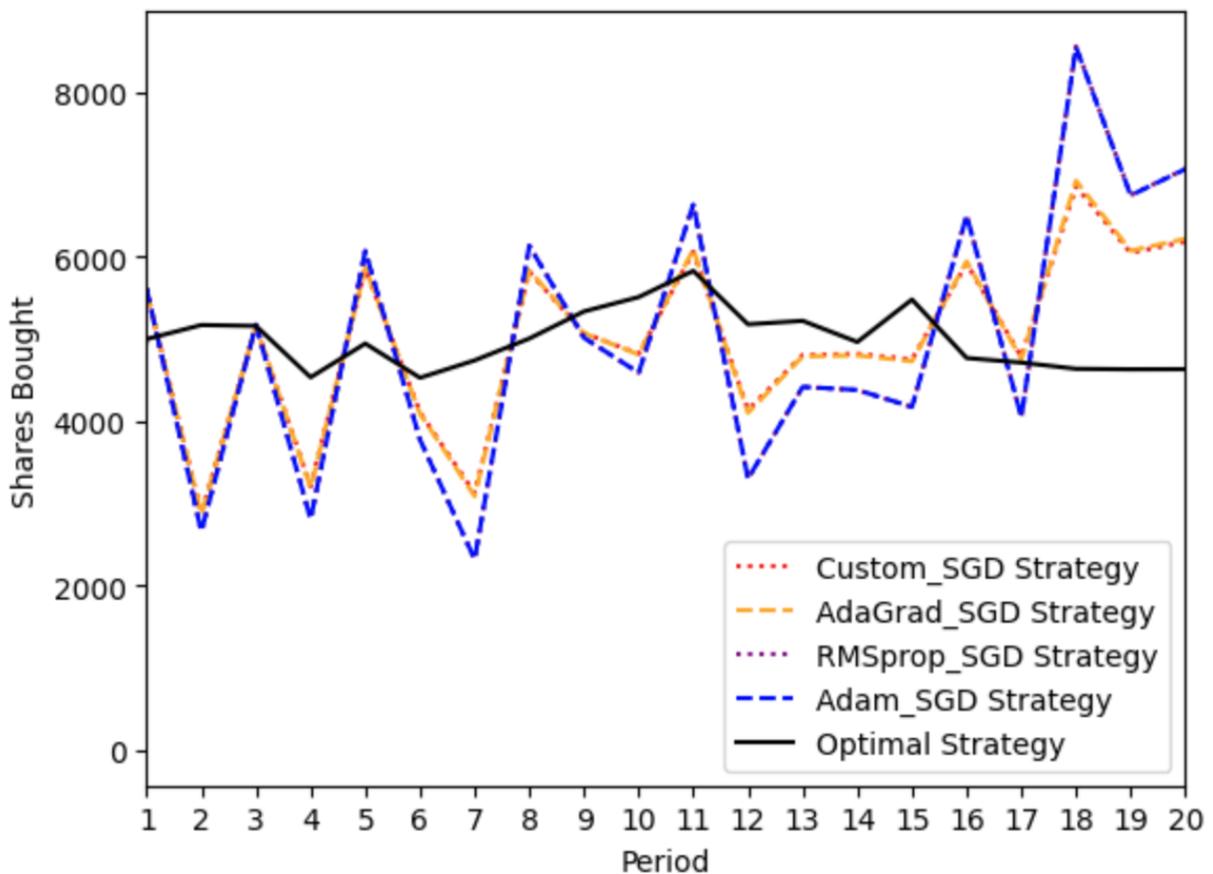

Figure 7 shows each strategy ranked by their simulated total execution costs:

Note that the RMSprop and Adam technique produced identical strategies, and the AdaGrad and Custom technique produced identical strategies.

```
          Execution Costs  Standard Deviation  Rank
Optimum       5.267080e+06         1118.689229     1
Custom        5.268126e+06         1487.208190     2
AdaGrad       5.268158e+06         1502.103699     3
Adam          5.269086e+06         1899.499653     4
RMSprop       5.269091e+06         1901.547911     5
```

Surprisingly the Custom and AdaGrad techniques, the simplest SGD algorithms performed the best at minimizing the accumulated execution costs and had a relatively low total standard deviation in $B_t$, in comparison to the other techniques. Additionally, the more dynamic NN oriented SGD algorithms, Adam and RMSprop, had the worst performance and the highest total standard deviation in $B_t$. However, it is likely that more meticulous hyper-tuning of model parameters needs to be conducted before making any conclusions. The Custom technique displayed the best performance, indicating that solving the implementation shortfall dilemma may not require such complex learning rate adjustment, and a standard ($\eta = 0.01$) with ($max\ i = 1,000$) may suffice. However, isolated analysis of these algorithms and their performance across the tuning of each individual hyper-parameter is required before making any conclusions about the true optima SGD model.

## Conclusion

In conclusion, even though none of the SGD algorithms were not able to beat the Optimum strategy derived by Bertsimas and Lo in the "Optimal Control of Execution Costs" (1998), the algorithms were able to derive alternative strategies that achiever relative converge to the true optimum solution, in practice. Though the SGD strategies displayed a significant increased variance in $B_t$, the algorithms still produced relatively similar total accumulated execution costs. Comparative analysis shows that the unique mathematical optimal is the true minimum execution cost, however, the SGD derived optimal $B_t$ and total execution costs are negligibly suboptimal. If, for example, an investor was not aware of the Optimum strategy, these results indicate it is possible to achieve relative optimality using stochastic gradient descent optimization techniques. These results may hold powerful implication, for example, the SGD approximation could provide insight and short-term solutions to mitigation implementation shortfall in markets that that evolve according to a complex non-linear price-impact dynamics. Specifically, for models whose mathematical optimum solutions have not yet been solved as an explicit function of market dynamics.